# QUALITY ASSURANCE AND INTEGRATION TESTING ASPECTS IN WEB BASED APPLICATIONS

Imran Akhtar Khan[1] and Roopa Singh[2]

[1]Dept. of Computer Engineering and IT, JJT University, Jhunjhunu, Rajasthan
`imran4bc@gmail.com`
[2] Dept. of Computer Engineering and IT, JJT University, Jhunjhunu, Rajasthan
`roopas1983@gmail.com`

## ABSTRACT

*Integration testing is one the important phase in software testing life cycle (STLC). With the fast growth of internet and web services, web-based applications are also growing rapidly and their importance and complexity is also increasing. Heterogeneous and diverse nature of distributed components, applications, along with their multi-platform support and cooperativeness make these applications more complex and swiftly increasing in their size. Quality assurance of these applications is becoming more crucial and important. Testing is one of the key processes to achieve and ensure the quality of these software or Web-based products. There are many testing challenges involved in Web-based applications. But most importantly integration is the most critical testing associated with Web-based applications. There are number of challenging factors involved in integration testing efforts. These factors have almost 70 percent to 80 percent impact on overall quality of Web-based applications. In software industry different kind of testing approaches are used by practitioners to solve the issues associated with integration which are due to ever increasing complexities of Web-based applications.*

## KEYWORDS

*(STLC) Software Development Life Cycle, Integration Testing, (SCM) Software Configuration Management, Web-based applications assurance.*

## 1. INTRODUCTION

Web-based applications typically work in a distributed, asynchronous fashion. These applications are very complex and their inter-dependency between different Web-components can cause more and more errors. The Web applications are accessed through the Web browser over internet or intranet. Since Web-based applications are distributed in nature, therefore it is not an easy task to test them. The uncovering of errors is very difficult in Web-based applications as compare to other traditional software applications. Testing of Web-based application is very difficult due to its nature like: heterogeneity, multi-platform support, autonomous, cooperative and distributed etc. Web-based applications are mostly complex software which are evolved and updated rapidly. Web applications have been integrated with mission critical systems by the different organizations due to which the quality and reliability of these applications is more and more crucial and therefore testing of these applications is very costly, time consuming and a big challenge. Web-based applications gather information and data from several heterogeneous sources. This raises the issue of integration as well.

The integration of different software components of Web-based applications is a taunting task. The usage of in-compatible technologies, differences in architecture of different components and applications can make it more difficult. Therefore, integration testing of Web applications is very crucial for the successful operations of these components among themselves. The integration testing of the components of Web-based applications involves a lot of factors, which might be raised during packaging, integration and deployment of the application. So, its integration testing is one of the challenging tasks. In Web applications the whole information with different structure and format is required to be integrated transparently and seamlessly [16].

## 2. WEB APPLICATIONS AND THEIR CHARACTERISTICS

A Web application is an application that is invoked with a client (mostly by Web browser) over the Internet, Intranet or Extranet. According to [24], A Web-based application allows the information processing functions to be initiated remotely from a client (browser) and executed partly on a Web server, application server and/or database server. These applications are specifically designed to be executed in a Web-based environment. When we visit on Web, we can find different kind of Websites. According to [23], in general, there are two types of Websites-
First one is based on the HTML (Hypertext Markup Language) also called static Websites and behave like simple printed newspapers or magazines. These Websites have published and printed materials for the end users. The examples of such kind of Websites are the different Websites of newspapers e.g. Hindustan Times, Times of India etc

Second one Websites enables the end users to interact with the Website. In this type Web pages are generated dynamically in the response of end user's input or action. These Websites work as software and utilities, also called as Web applications. Web applications run on servers and end users access these applications through Web browsers [23]. The examples of Web applications are supply chain management, online banking systems, online retail systems and different email services like Google, yahoo and hotmail.

Every software application has their specific characteristics, some of those are common and some are related to specific applications. These characteristics are very important to keep in mind before the development and testing of any application. The following general characteristics and attributes are encountered in majority of Web-based applications:

- **Network intensive:** Since Web applications are delivered to a diverse community of users, the nature of Web-based applications is network intensive. These applications reside on network (like Internet, Intranet or Extranet).
- **Content:** Web applications are heavily content-driven because mostly Web applications present textual description, graphical data, audio and video or multi-media information to the end users.
- **Continuous evolution:** Most Web applications evolve continuously. These applications updated on the regular interval, even some applications are updated on hourly schedule to provide latest information to the end users.
- **Short development schedule:** Mostly Web-based applications have very tight development schedule. It means these applications have very short time for the development and developed under compressed time schedule. The time to market for a complete Website from planning to implementation and testing can be a matter of few days or weeks.
- **Security:** To protect sensitive content and information provided by the user, and for successful data transmission strong security measures are implemented in the Web-based applications.

- **Aesthetic:** One of the most important characteristics of Web applications is aesthetic appearance. Aesthetic appearance of those Web-based applications which designed for selling products and ideas is as important as technical design. The above all are some simple but important characteristics of Web-based applications, but as the complexities of these applications are growing, some other characteristics such as distributed, heterogeneity, autonomous, dynamic, hypermedia, multiplatform support, ubiquitous are very important to understand. These characteristics are the important factors that are necessary to keep in mind before the designing, implementing, testing and deployment of the Web-based applications. These characteristics can assist the developers and software engineers to built successful applications.

## 3. WHY TESTING WEB APPLICATIONS IS DIFFERENT?

Testing web applications is different because of many factors, scenarios affecting the performance and user experience. Web applications can typically cater to a large and a diverse audience. Web Applications can also be exposed to wide range of security threats. Web applications may open up illegal points of entry to the databases and other systems holding sensitive information. To ensure that the web application works reliably and correctly under different situations these factors need to be accounted for and tested. Hence a lot of effort needs to put in for Test Planning and Test Design. Test Cases should be written covering the different scenarios not only of functional usage but also technical considerations such as Network speeds, Screen Resolution, etc. For example an application may work fine on Broad Band internet users but may perform miserably for users with dial up internet connections. Web Applications are known to give errors on slow networks, whereas they perform well on high speed connections. Web pages don't render correctly for certain situations but work okay with others. Images may take longer to download for slower networks and the end user perception of the application may not be good.

## 4. QUALITY ASSURANCE OF WEB-BASED APPLICATIONS

The success of software engineering depends upon the delivery of high quality software. Quality is one of the key factors in the market growth and success of a product. In recent years, quality of software product and quality in service has become principles for many corporations and organizations to distinguish themselves from competitors and to cover larger market place. Quality is an ambiguous word; means there are lot of definitions available for quality. According to IEEE, quality is the degree to which software meets customer or user needs or expectations. The simplest definition of quality is in the mean of customer satisfaction. Robert Glass [28] summarize the customer satisfaction in mathematical equation as –

"*Customer Satisfaction = Compliant Product + Good Quality + Delivery within Schedule and Budget*"

He also argues that quality is an important factor in the development of a product, but if the customer is not satisfied then nothing else really matters. Quality assurance includes all the process-related activities to achieve the quality. It is involved from the start of a project. In other words we can say that it is an umbrella activity which is applied on each step in the software development process. It controls the insight and outsight quality of the software. According to [23][27], Software quality assurance include the following important elements.

International Journal of Computer Science, Engineering and Applications (IJCSEA) Vol.2, No.3, June 2012- ✓ Quality management tools.
- ✓ Effective software engineering methods and tools.
- ✓ Formal technical reviews applicable to whole software process.
- ✓ Effective testing strategies and techniques.
- ✓ Procedures to control documentations and changes in it.
- ✓ Procedures to assure compliance to standards.
- ✓ Mechanisms for measurement and reporting.

The dependency of people on Web applications is increasing continuously due to which the Web systems have become more and more complex. These applications are increasingly integrated in business strategies of small and large organizations. Therefore quality, reliability, accessibility, usability, adaptability and functionality have become very crucial and important factors for the Web applications. The process of Web engineering is used to develop the high quality Web applications [23]. It defines the specific techniques, methodologies and models to develop Web-based applications. The main aim of engineering of Web-based applications is to attain and produce high quality software products. Quality assurance of Web-based applications is very crucial and vital to achieve the high quality. The quality assurance of Web applications is the responsibility of Web developers and Software quality assurance group. To ensure the quality of Web applications the development team should follow the above mentioned methods of software quality assurance [23]. That is, Web developers should follow the quality management approach, effective software methods and tools, formal reviews, effective testing strategies and techniques, follow the standards and usage of appropriate mechanisms for measurement and reporting.

### 4.1 QUALITY ATTRIBUTES OF WEB-BASED APPLICATIONS

Different persons (end-users of Web applications) have different views and opinions about the good Web Application. These opinions and views depend upon the end user and vary widely, because some individuals like flashy graphics and some want simple text. Some users want detailed information and some only like short and abbreviated presentations. It is fact that the user perception of likeness of Web application might be more important that any technical discussion of Web applications quality. This raise the question about the perception of quality of Web application and about the different attributes that must be exhibit to achieve goodness in the eyes of the end-users and also exhibit the technical characteristics of quality that enable the Web engineers to enhance, adapt, correct and support the Web application over the long term. [23] Almost all general quality characteristics can be applied to Web applications but the most important and relevant quality attributes are prepared by [27], who developed a quality requirement tree that identifies a set of attributes that lead to develop high quality Web applications.

### 4.2 QUALITY ASSURANCE ENABLING TECHNOLOGIES

In order to build reliable and high quality Web applications the Web engineer should be familiar with the quality assurance enabling technologies. These enabling technologies are component based development, internet standards and security [23]. The brief description of these enabling technologies is as under:

- **Component based Development:** The explosive growth of Web-based applications has evolved the component technologies. The available famous infrastructure standards for web development are CORBA (Common Object Request Broker Architecture), COM/DCOM (Component Object Model/ Distributed Component Object Model) and JavaBeans. These

112

different standards are helpful to deploy and integrate third party components and to develop custom components to communicate with each other and with other system services.

- **Internet Standards:** Internet standards are specifications which are stable and well-understood, has multiple, independent, and interoperable implementations with operational experience and recognizably useful in some or all elements of the Internet. In early 1990's HTML (Hyper Text Markup Language) was the dominant standard to develop Web applications but now the Web applications have become more complex, their size is growing, therefore new standards have emerged. XML (Extensible Markup Language) and XHTML (Extensible Hypertext Markup Language) are new standards that have been adopted to develop next generation Web applications. These standards allow developers to define their own custom tags to describe the content of Web pages in Web applications. By following these standards the quality of Web applications is increased in the mean of robustness, interoperability, integration, functionality, reliability and accessibility.

- **Security:** When our application deployed and launched on the network or Internet, there are great risks of unauthorized used. There are great threats of vulnerabilities. The hackers try to unauthorized access in the intent of some profit or for some other aims. Sometimes internal personnel can be involved in unauthorized access of particular application for their specific benefits and aims or malicious intents. Therefore security measures are very important to build high quality Web applications. A lot of security measures are being applied to minimize the threats of vulnerabilities and malicious use of the particular applications like firewall, encryption, and other security policies.

## 5. PURPOSE OF INTEGRATION TESTING

The purpose of integration testing is to make sure that modules and their interfaces in an application interact with each other in a correct and secure way. Basically integration testing is based on functional requirements specification and design which are used as an input in integration testing process. Theoretically many integration testing techniques are available but some of them only give the proper guidelines to make correct test cases such as bottom-up, top-down, black box, sandwich, incremental and big-bang. Integration testing covers following types of concerning areas during integrating different modules:

- Calls of different software components, while interacting to each other
- Data and information sharing between the modules in proper manners
- Compatibility, which ensures one module that does not effect on the performance and functionality of the other modules.
- Non-functional issues

It is important that integration testing should be conducted in development environment or separate testing environment instead of live environment. Most of the times, integration testing is done by the development team supervised by development team leader and a person from quality assurance group. In such circumstances the responsibility of development team leader is to ensure that integration testing must be progressed with suitable choice of test techniques. The development team leader then provides results of testing to the test team leader.

## 6. INTEGRATION TESTING CHALLENGES

The size of Web applications are growing due to involvement of new emerging process like business processes and highly secure requirements from customers. The existing integrated solutions are the cheapest and fast way to develop such kind of large Web-based applications. But testing of such application is complex task due to its large size, integration of multilingual components and use of different operating systems. The most important and commonly known challenges during integration testing are: [1]

- **Heterogeneous Infrastructure and Environment:** Web-based applications run on different environment like different browsers and may have number of different heterogeneous components. It means, heterogeneity is one of the key features of Web-based applications. According to [1], heterogeneity may introduce the incompatibilities between different programming languages, databases, different operating systems and external operational environments involved in development and deployment of Web-based applications. Complex integrated solutions consist upon different software components which are developed by using different programming languages and technologies. The assurance of compatibility and interoperability between these components is one of major concern during the testing process.

- **Service Oriented Environment:** In Service Oriented environment, it is very difficult to diagnose the defects for integration testing team because data is encapsulated in the messages in transport protocols. Therefore it is very difficult to find out the error until complete system is fully implemented. Web services and SOA demand very explicit inputs and outputs. In this environment many applications sends only update or received messages, there is no guarantee that data came from system A to system B is accurate.

- **Heterogeneous Database:** Databases are the most important components of Web-based applications for the management and flow of data. Different types and versions of databases are available in the market such as MySQL, SQL Server and DB2 Server etc, so there is always chance of incompatibility issues among the different databases and software components that access these databases. The assurance of compatibility between different software, operating systems and operational environments are also very critical tasks. According to [29], various semantic conflicts occurs among heterogeneous data cubes, for example value to value conflicts, data representation conflict, data scaling conflicts, Inconsistent data, value to attribute conflicts, value to table conflicts, attribute to attribute conflicts, attribute to table conflicts.

- **Inconsistent Interaction Models:** Web-based applications are based on number of different Web components which are developed by different group of teams by using different methods and approaches. According to [1], in complex Web-based applications, control protocols and data models play key role in the reliable communication and interaction among different integrated subsystems. Control protocols are responsible of defining rules that how integrated components interact to each other. Data models define the contents and format of communication between them. Since Web-based applications can be based on number of different components and most of the time different groups of developers are involved in development process. They may have different views and assumptions about the development of those components and their interaction (i.e. control protocols and data models) with each other. Except these constraints; different components of Web-based applications may expose multiple interfaces that can vary these constraints and also change the type of the relationship between the components.

- **Distributed Nature of Systems:** Web-based applications are mostly developed under distributed environments, so the issues related to distributed systems such as race condition and dead lock can be inherited [1]. Distributed nature of systems can have great impact on working of Web-based applications and these issues can be solved during integration testing. Existence of more than one versions of same software component generates multi version issues in system.

## 7. CONCLUSION

Web based applications are playing very important role in many business domain like retail, finance, sales, marketing and management. Software integration also can not be neglected in large organization and critical software application. Since integration testing usually runs through much iteration and implemented system is tested again and again with the interacting component because more testing leads finding of more bugs. Therefore it is very important that good software management policy should be in place. We should be able to track the components and their versions. So each time we integrate the application components we know exactly what versions go into the build process. There may be the chance that wrong builds were sent or wrong version of build were sent or there are some missing components and hence a lot of manual effort will be required. There should be proper build process and we should have well written script to integrate and deploy the components. If the defects are not logged correctly then integration testing may loose the track. Each defect should be logged in the defect tracking tool for example HP Mercury Tool. In this tool information should be captured who is the owner of the defect, who will resolve it, what is the description of the defect and what is the final status like defect is closed or in open status.

## Authors

**Imran Akhtar Khan**

His research area is Software Testing- especially skilled in Integration testing, Web based testing. He has completed his Master of Computer Science and Applications (MCA) degree from Aligarh Muslim University (AMU), Aligarh.

**Roopa Singh**

Her area of interest is – E-Commerce (B2C and B2B), Web based System and Integration testing. She has received her Master of Computer Science and Applications (MCA) degree from Aligarh Muslim University (AMU), Aligarh.